\magnification=1200
\baselineskip=18truept
\input epsf


\def\draftversion{N}

\if \draftversion Y


\fi

\rightline{DPNU-97-08}
\rightline{March 1997}
\medskip

\vskip 2truecm
\centerline{\bf Four-Dimensional ${\cal N}=1$ Supersymmetric Yang-Mills Theory}
\centerline{\bf on the Lattice without Fine-Tuning}
\vskip 1truecm
\centerline{Jun Nishimura}
\centerline {Department of Physics, Nagoya University}
\centerline {Chikusa-ku, Nagoya 464-01, Japan}

\vskip 1truecm
\centerline{e-mail: {\tt nisimura@eken.phys.nagoya-u.ac.jp}}

\vfill
\centerline{\bf Abstract}
\vskip 0.75truecm
We propose a method to formulate four-dimensional ${\cal N}=1$
super Yang-Mills theory on the lattice without fine-tuning.
We first show that four-dimensional 
Weyl fermion in a real representation, 
which is equivalent to Majorana fermion, 
can be formulated using the domain wall approach with an addition of 
a Majorana mass term only for the unwanted mirror fermion.
This formalism has manifest gauge invariance.
Fermion number conservation is violated only by the additional
Majorana mass term for the mirror fermion and the violation 
is propagated to the physical fermion sector through anomalous currents.
Due to this feature, the formalism, when applied to the gluino in the present
case, ensures the restoration of supersymmetry in the continuum limit
without fine-tuning, unlike the proposal by Curci and Veneziano.
\bigskip
\leftline{PACS: 11.15.Ha; 11.30.Pb}
\leftline{Keywords: Lattice gauge theory, Supersymmetry}
\medskip

\vfill\eject

Supersymmetry is one of the most exciting topics in field theory.
It is important in two respects.
In phenomenology, supersymmetry is motivated as a natural solution 
to the gauge hierarchy.
From purely field theoretical points of view, supersymmetry enables
analytic study of nonperturbative aspects of field theories.
On the other hand, lattice formalism has been a powerful tool to 
extract nonperturbative dynamics of field theories.
It would be nice to use this formalism to explore the nonperturbative 
dynamics of supersymmetric theories.

In spite of much effort in this direction, most of the attempts 
seem to have failed in practice.
A practical proposal, however,
has been given by Curci and Veneziano [1].
They propose to give up manifest supersymmetry on the lattice,
and instead, to restore it in the continuum limit.
It has been shown that this can indeed be done
for four-dimensional ${\cal N}=1$ super Yang-Mills theory
using the Wilson-Majorana fermion for the gluino
and fine-tuning the hopping parameter to the chiral limit.

Using this proposal, some numerical simulations have been started [2].
Although recent developments in treating dynamical fermions help
quite a lot,
it is needless to say that it would be better if we could do without
fine-tuning.
Instead of fine-tuning a bare parameter to the chiral limit,
we could impose chiral symmetry on the lattice theory.
This is not easy in standard lattice formulations of fermions, though [3].
In the section 9 of Ref.~[4], 
it has been suggested that the overlap formalism can be used
as such a formulation.
The method we propose provides an alternative way, which is
more suited for numerical study.

There is a formulation [5,6] that preserves chiral symmetry
for Dirac fermion using the domain wall formalism [7].
We would like to propose a Majorana version of it.
What we consider directly is Weyl fermion in a real representation, 
which is equivalent to Majorana fermion in four dimensions.

In Ref.~[8], a proposal for formulating 
anomaly-free chiral gauge theories has been given.
Their idea can be summarized as follows.
Try to formulate it following Kaplan's proposal [7] for formulating
chiral gauge theory on the lattice.
We consider an extra dimension and consider five-dimensional 
Dirac fermion, whose mass has a domain-wall type dependence on
the fifth coordinate.
On the domain wall, we have a chiral zero mode, which could be used to
construct chiral gauge theory.
When the extent in the extra dimension is finite, however,
we encounter an anti domain wall on which
an unwanted mirror fermion with opposite chirality appears.
Their idea to remedy this situation is 
to add a four-fermi gauge-invariant term 
motivated by Eichten-Preskill [9],
only for the mirror fermion 
in order to give it a mass of the order of the cutoff,
and thus to make it decouple in the continuum limit.

We apply this idea to the present case.
Now since the Weyl fermion is in a real representation,
we can introduce a gauge-invariant Majorana mass term only 
for the mirror fermion that serves for the same purpose.
Here the dynamical effect of the additional term is quite clear,
as compared with the four-fermi interaction in Ref.~[8],
and therefore it does not raise any further subtle problems.
Fermion number violation in the physical fermion sector
comes only from anomalous currents which pick up the violation
due to the additional Majorana mass term for the mirror fermion.
Restating this feature in terms of Majorana fermion, which is equivalent
to Weyl fermion in a real representation,
we have chiral symmetry up to the anomaly.
Using this formalism for the gluino, we can obtain
four-dimensional ${\cal N}=1$ super Yang-Mills theory in the continuum limit
without fine-tuning.



\bigskip

Let us define the model we propose.
We consider five-dimensional Dirac fermion on the lattice
using the Wilson fermion formalism [10].
We take the coordinate in the fifth direction to be 
$-L_5 \le x_5 \le L_5$.
The boundary condition is taken to be periodic in the fifth direction
and either periodic or anti-periodic in the other four directions.
The action is given by
$$S =  \sum_x 
[ \bar{\Psi}(x) \gamma_a D_a  \Psi(x) 
+m(x_5) \bar{\Psi}(x)  \Psi(x) 
+ {1 \over 2} \bar{\Psi}(x) \triangle \Psi(x) ],
\eqno{(1)}$$
where $\Psi(x)$ is a four-component spinor,
and $\gamma_a~(a=1,\cdots,5)$ are the gamma matrices in five dimensions.
The mass $m(x_5)$ has the domain-wall type dependence on $x_5$:
$m(x_5)=m_0$ for $1 \le x_5 \le L_5$ and
$m(x_5)=-m_0$ for $-L_5+1 \le x_5 \le 0$.
The $D_a$ and $\triangle$ are defined as follows.
$$ D_a = {1\over 2}  (\nabla_a^\ast + \nabla_a), \eqno{(2)}$$
$$ \triangle = \nabla_a^\ast \nabla_a, 
\eqno{(3)}$$
where 
$$
\nabla_a \Psi(x) = U_a(x) \Psi(x+\hat{a})-\Psi(x),
$$
$$
\nabla_a^\ast \Psi(x) = \Psi(x)- U_a^\dagger(x-\hat{a}) \Psi(x-\hat{a}).
\eqno{(4)}$$
The system is invariant under the gauge transformation.
$$
\eqalign{
\Psi(x) & \rightarrow g(x)~\Psi(x), \cr
\bar{\Psi}(x) & \rightarrow \bar{\Psi}(x)~ g^\dagger (x), \cr
U_a(x)  & \rightarrow g(x)~ U_a(x) ~ g^\dagger (x+\hat{a}). \cr}
\eqno{(5)}$$
The link variables $U_a(x)$ are taken such that
$U_5(x)=I$ and $U_\mu(x)~(\mu=1,\cdots,4)$ are independent 
of $x_5$.

In the following we take the following representation for the
gamma matrices.
$$\gamma_\mu = \pmatrix{0 & \bar{\sigma}_\mu  \cr 
\sigma_\mu & 0 \cr}~~~~for~\mu=1,\cdots,4;\ \ \ \
\gamma_5 = \pmatrix{I & 0 \cr 0 & -I \cr},\eqno{(6)}$$
where $\sigma_\mu = (1,i\sigma_i)$,
$\bar{\sigma}_\mu = (1,-i\sigma_i)$ 
and $\sigma_i$ are the Pauli matrices.
Now we decompose the four-component spinor $\Psi(x)$ as
$$\Psi=\pmatrix{\psi \cr  \chi \cr};\ \ \ \
\bar{\Psi}=\pmatrix{ \bar{\psi} & \bar{\chi} \cr} \gamma_0 =
\pmatrix{ \bar{\chi} & \bar{\psi} \cr} .\eqno{(7)}$$
We obtain 
$$
\eqalign{
S & =  \sum_x 
[\bar{\psi}(x)  \sigma_\mu D_\mu  \psi(x) 
+ \bar{\chi}(x) \bar{\sigma}_\mu D_\mu  \chi(x)  \cr
& ~~~ + \bar{\chi}(x) D_5 \psi(x) - \bar{\psi}(x) D_5 \chi(x)  \cr
& ~~~ +m(x_5) (\bar{\chi}(x)  \psi(x) + \bar{\psi}(x)  \chi(x) )
+ {1 \over 2}
(\bar{\chi}(x) \triangle \psi(x) + \bar{\psi}(x) \triangle \chi(x) )].
\cr}
\eqno{(8)}$$
Due to the argument in Ref.~[7],
we have a chiral zero mode in $\psi(x)$ 
localized around the first wall $x_5=0$, 
while we have another chiral zero mode in $\chi(x)$ 
localized around the second wall $x_5=L_5$ .
There are many other massive modes, whose contribution has to be 
appropriately canceled by additional five-dimensional massive scalar fields
as in [5], which we omit in this paper.
We can see that the theory has exact chiral symmetry in the $L_5 
\rightarrow \infty$ limit by rewriting it in terms of the
overlap formula using the transfer matrix formalism
as has been done in Ref. [11].
$|0 \pm \rangle$ are defined as
the ground states of the transfer matrices $\hat{T}_\pm$ in the fifth
direction,
whose explicit forms are given by eq.(2.12) of Ref. [11].
$\pm$ correspond to the positive $m$ region
and the negative $m$ region, respectively.
The fermion determinant can be expressed as
$\langle 0 - |0 + \rangle  \langle 0 + | 0 - \rangle$.
The two overlaps correspond to the chiral fermions
on the two walls.
When the gauge configuration is topologically trivial,
$|0 \pm \rangle$ have the same fermion number and the fermion number
is conserved separately on the two walls.
On the other hand, when the gauge configuration is topologically non-trivial,
$|0 \pm \rangle$ have different fermion numbers and the fermion number
is violated on each wall, though conserved as a whole.
Thus we have exact chiral symmetry up to the anomaly.
Note that the symmetry cannot be seen in the action (8).
This is because the $L_5 \rightarrow \infty $ limit 
is essential in obtaining the exact symmetry.
When $L_5$ is finite, the chiral symmetry is broken due to the
non-vanishing overlap of the wave-functions of the chiral zero modes
on the two walls.
However, this is expected to be exponentially small 
for sufficiently large $L_5$ and thus we have almost exact chiral symmetry.

Now let us consider the chiral zero mode on the first wall $x_5=0$ 
as the physical Weyl fermion and 
the one on the second wall $x_5=L_5$ as the unwanted mirror fermion.
In order to give the mirror fermion a mass,
we add a Majorana mass term to the action localized on the second wall.
$$
S_{add} = \left. M \sum_{x} [ \chi^T(x)\sigma_2 \chi(x) + 
\bar{\chi}(x)\sigma_2 \bar{\chi}^T(x) ]  \right|_{x_5=L_5},
\eqno{(9)}$$
where $M$ is kept fixed when one takes the continuum limit.
Note that this term is gauge invariant when 
the fermion is in a real representation of the gauge group, since
we have $g^T(x) ~ g(x) = 1$.
Note also that it is invariant under four-dimensional rotation and 
translation.
Let us see what we obtain in the $L_5 \rightarrow \infty$ limit.
Now the fermion determinant can be written as
$\langle 0 - |0 + \rangle  \langle 0 + |{\cal O}| 0 - \rangle$,
where ${\cal O}$ is an operator which violates fermion number.
Hence the chiral fermion on the second wall acquires Majorana mass
of the order of the cutoff,
while the chiral fermion on the first wall remains massless.
The fermion number of the chiral fermion on the first wall is conserved
up to the anomaly.
The violation of the symmetry when the $L_5$ is finite is expected
to be exponentially small as before.

Thus we can formulate four-dimensional chiral gauge theory with
Weyl fermion in a real representation
using the domain wall approach with an addition of the Majorana
mass term for the unwanted mirror fermion.
This is not so surprising, because Weyl fermion in a real representation
in four dimensions is equivalent to Majorana fermion and
thus the theory is essentially vector-like.
However, the important feature of this formalism, 
as compared with the Wilson fermion formalism for the Majorana fermion,
is that the theory has the chiral symmetry up to the anomalous currents
propagating from the mirror fermion sector.
Due to this feature,
we can use the formalism for a gauge theory with massless Majorana fermion
in the adjoint representation, which is expected to be supersymmetric
in the continuum limit without fine-tuning.

In the large $N$ limit of the gauge group, 
there is yet another possibility to deal with
supersymmetry as was suggested in Ref.~[12].
We might be able to use the continuum version of 
the Eguchi-Kawai model.
The advantage of this approach is that it has manifest supersymmetry.
Indeed there is a revived interest in this model in the context of
superstring theory [13].
In contrast to the lattice version of the Eguchi-Kawai model [14], which
is known to be equivalent to the infinite volume lattice theory,
in the continuum version, the connection to the 
infinite volume continuum theory is not so obvious beyond perturbation 
theory.
It is therefore worth while to study the large $N$ limit of 
super Yang-Mills theory using our method.
For this purpose, the reduction of dynamical degrees of freedom
in the large $N$ limit [14] is essential.
Since our formalism is an ordinary Lagrangian field theory 
with all the fields in the adjoint representation,
we can perform the twisted reduction procedure [15] for the 
four-dimensional space-time directions.
Thus, our model in the large $N$ limit 
is equivalent to the reduced model in which the space-time 
direction is reduced, while the extra direction is left unreduced.
When one uses the overlap formalism [4] to deal with super Yang-Mills
theories without fine-tuning, the large $N$ reduction is not obvious.
This is another advantage of the present approach.

\bigskip
\medskip

To summarize, we have proposed a method to deal with
four-dimensional ${\cal N}=1$ super Yang-Mills theory on the lattice
without fine-tuning.
A formalism for dealing with Majorana fermion with
chiral symmetry up to the anomaly has been given.
Using this formalism for the gluino, the supersymmetry is expected 
to be restored in the continuum limit without fine-tuning,
which makes studies in this direction much more efficient.
One of the most interesting physics accessible with the present method
is the gluino condensation.


\bigskip
\centerline{\bf Acknowledgement}
\medskip

The author would like to thank R. Narayanan and M. Oshikawa
for helpful discussions.
He is also grateful to D.B. Kaplan for valuable comments.


\vfill\eject
\centerline{\bf References}

\bigskip

\item{1.} G. Curci and G. Veneziano, Nucl. Phys. B292 (1987) 555.
\item{2.} I. Montvay, Nucl. Phys. B466 (1996)  259;
A. Donini and M. Guagnelli, Phys. Lett. B383 (1996) 301.
\item{3.} H. Nielsen and M. Ninomiya, Nucl. Phys. B185 (1981) 20;
Nucl. Phys. B193 (1981) 173; B195 (1982) 541(E).
\item{4.} R. Narayanan and H. Neuberger, Nucl. Phys. B443 (1995) 305.
\item{5.} V. Furman and Y. Shamir, Nucl. Phys. B439, 54 (1995).
\item{6.} T. Blum and A. Soni, preprint hep-lat/9611030 (1996).
\item{7.} D. Kaplan, Phys. Lett. B288 (1992) 342;
M. Golterman, K. Jansen and D. Kaplan, Phys. Lett. B301 (1993) 219.
\item{8.} M. Creutz, M. Tytgat, C. Rebbi and S.-S. Xue,
preprint hep-lat/9612017 (1996).
\item{9.} E. Eichten and J. Preskill, Nucl. Phys. B268 (1986) 179.
\item{10.} K. Wilson, in {\sl New Phenomena in Subnuclear Physics},
Edited by A. Zichichi (Plenum Press, NY, 1977).
\item{11.} R. Narayanan and H. Neuberger, Nucl. Phys. B412 (1994) 574.
\item{12.} D. Gross and Y. Kitazawa, Nucl. Phys. B206 (1982) 440.
\item{13.} T. Banks, W. Fischler, S. Shenker and L. Susskind, preprint,
hep-th/9610043 (1996); 
N. Ishibashi, H. Kawai, Y. Kitazawa and A. Tsuchiya,
preprint, hep-th/9612115 (1996).
\item{14.} T. Eguchi and H. Kawai, Phys. Rev. Lett. 48 (1982) 1063.
\item{15.} A. Gonzalez-Arroyo and M. Okawa, Phys. Rev.
D27 (1983) 2397;
T. Eguchi and R. Nakayama, Phys. Lett. 122B (1983) 59.

\end